\documentclass[aps,amsmath,notitlepage,twocolumn,amssymb,prb,letterpaper,longbibliography]{revtex4-1}

\usepackage{graphicx}
\usepackage{dcolumn}
\usepackage{bm}
\usepackage[colorlinks=true,citecolor=red,urlcolor=blue]{hyperref}
\usepackage{wasysym}
\usepackage{stmaryrd}
\usepackage{verbatim}

\usepackage{epstopdf}
\usepackage{subfigure}
\usepackage{tabularx}
\usepackage{amsfonts}
\usepackage{wasysym}
\usepackage{bm}

\newcommand{\be}{\begin{equation}}
\newcommand{\bea}{\begin{eqnarray}}
\newcommand{\ee}{\end{equation}}
\newcommand{\eea}{\end{eqnarray}}

\newcommand{\vev}[1]{\left\langle#1\right\rangle}

\begin{document}
\title{Kitaev materials beyond iridates: order by quantum disorder and Weyl magnons 
in rare-earth double perovskites}
\author{Fei-Ye Li$^{1}$}
\thanks{These two authors contribute equally.}
\author{Yao-Dong Li$^{2}$}
\thanks{These two authors contribute equally.}
\author{Yue Yu$^{2,4}$}
\author{Arun Paramekanti$^{3}$}
\author{Gang Chen$^{2,4,5}$}
\email{gangchen.physics@gmail.com, gchen$_$physics@fudan.edu.cn}
\affiliation {${}^1$CAS Key Laboratory of Theoretical Physics, 
Institute of Theoretical Physics, Chinese Academy of Sciences, 
Beijing 100190, People's Republic of China}
\affiliation{${}^2$State Key Laboratory of Surface Physics, 
Center for Field Theory and Particle Physics, 
Department of Physics, Fudan University, Shanghai 200433, People's Republic of China}
\affiliation{${}^3$Department of Physics, University of Toronto, Toronto, Ontario, Canada M5S 1A7 and Canadian Institute for Advanced Research, Toronto, Ontario, Canada M5G 1Z8}
\affiliation{${}^4$Collaborative Innovation Center of Advanced Microstructures, 
Nanjing, 210093, People's Republic of China}
\affiliation{${}^5$Perimeter Institute for Theoretical Physics, Waterloo, Ontario N2L 2Y5, Canada}
\date{\today}

\begin{abstract}
Motivated by the experiments on the rare-earth double perovskites, 
we propose a generalized Kitaev-Heisenberg model to describe the generic
interaction between the spin-orbit-entangled Kramers’ doublets of the 
rare-earth moments. We carry out a systematic analysis of the mean-field 
phase diagram of this new model. In the phase diagram, there exist 
large regions with a continuous $U(1)$ or $O(3)$ degeneracy. 
Since no symmetry of the model protects such a continuous degeneracy,
we predict that the quantum fluctuation lifts the continuous 
degeneracy and favors various magnetic orders in the phase diagram.  
From this order by quantum disorder mechanism, we further predict that 
the magnetic excitations of the resulting ordered phases are 
characterized by nearly gapless pseudo-Goldstone modes. 
We find that there exist Weyl magnon excitations for certain magnetic 
orders. We expect our prediction to inspire further study of Kitaev physics, 
the order by quantum disorder phenomenon and topological spin wave modes 
in the rare-earth magnets and the systems alike.
\end{abstract}

\maketitle

\section{Introduction}
\label{sec1}

There has been an intensive interest in the study of Kitaev materials~\cite{Jackeli2009,Jackeli2010,Hongchen2011,Mei2012,Mazin2012,
PhysRevB.89.014414,Chaloupka2013,Simon2015,
Takagi2015,Perkins2015,Modic2014,Nishimoto2016,Rau2014,Lou2015,EricLee2016}. 
Originally, Kitaev materials refer to
honeycomb~\cite{Jackeli2009,Jackeli2010,Singh2012,Choi2012}, 
hyperhoneycomb~\cite{Takagi2015,lee2014order,PhysRevB.90.134425}, 
harmonic honeycomb~\cite{Modic2014}, 
and hyperkagome iridates~\cite{Okamoto2007,Chen2008,Chen2013}, 
and more recently, have been extended to the new material 
RuCl$_3$~\cite{Sears2015,Kim2015,Banerjee2016}. In these systems, 
the magnetic ions are heavy elements like Ir$^{4+}$ and Ru$^{4+}$, 
where the spin-orbit coupling (SOC) is quite strong. 
Due to the spin-orbit entanglement of the local moments, 
the interaction between them depends on the bond
orientation~\cite{Chen2008,Jackeli2009,Chen2010,Chen2011}, 
and may involve a large Kitaev spin interaction~\cite{Jackeli2009}. 
Since Kitaev model~\cite{Kitaev20062} supports a robust 
quantum spin liquid ground state, one goal of exploring these systems 
is to realize the Kitaev spin liquid with a dominant Kitaev interaction. 
More generally, it is of great importance to understand the role of 
spin-orbit entanglement on the properties of a strongly 
correlated quantum many-body system~\cite{Chen2014}. 

Since the Kitaev interaction~\cite{Jackeli2009}, or more precisely, 
the bond dependent spin interaction, 
is a natural consequence of the strong SOC~\cite{Chen2008}, 
its presence should go beyond iridates or ruthenates. 
The vast families of rare-earth magnets have {\sl never} 
been explored along the line of Kitaev interaction. In fact, rare-earth moments
have much stronger SOC than iridium or ruthenium~\cite{Curnoe2008,RevModPhys.82.53,Onoda2011,Sungbin2012,Huang2014,Yaodong2016,Savary2013,li2015rare,li2016anisotropic}.
The $4f$ electrons are much more localized than the $5d$ or $4d$ electrons
in iridates and ruthenates. Most often, the interaction between the local moments
in the rare earth systems is merely restricted to the nearest neighbors,
while the iridates or ruthenates may involve significant further neighbor
interactions due to the extended electron wavefunctions~\cite{pesin2010mott}. 
Moreover, the rare-earth elements do not suffer from the neutron absorption
issue that prevails in the study of iridates~\cite{Singh2012,Choi2012}. 
Because of the small energy scale of the interaction, the external magnetic field
and the inelastic neutron scattering could even be used to precisely 
determine the Hamiltonian of the rare-earth systems. All these advantages 
make the rare-earth systems ideal Kitaev materials.

\begin{figure}
\includegraphics[width=0.3\textwidth]{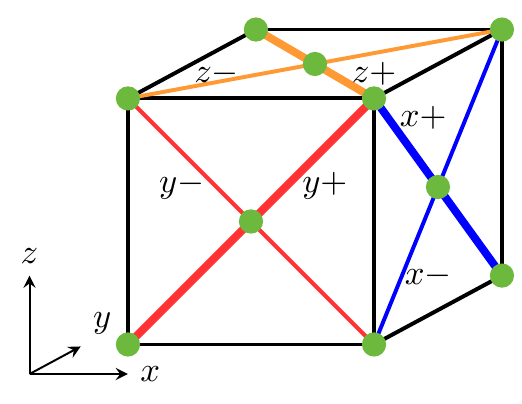}
\caption{(Color online.) The bond dependent interactions in the FCC lattice. 
We have marked the six distinct bond types $\gamma \pm$ ($\gamma=x, y, z$), 
that have the specific forms of bond-dependent interactions in Eq.~\ref{eq: ham}. 
The inset is the global coordinate system that defines the spin components. 
	}
\label{Fig1}
\end{figure}

In this paper, we turn from iridates to the rare-earth systems 
and explore the consequence of the spin-orbit entanglement and the 
Kitaev interaction in rare-earth double perovskites. 
Double perovskite (A$_2$BB$'$O$_6$) is a very common system in 
which the magnetic ions B$'$ form a face-centered-cubic     
(FCC) lattice~\cite{Chen2010,Chen2011,Shumpei2015}. 
Previously, the interplay of strong correlation 
and strong SOC has been explored for the $4d$ and $5d$ transition
metal elements with partially filled $t_{\text{2g}}$ 
shells~\cite{Chen2010,Chen2011}. It was pointed out that the 
strong spin-orbit entanglement gives a multipolar structure 
of the local moments and rich magnetic multipolar orders~\cite{Chen2010,Chen2011}. 
In contrast, the rare-earth electrons often experience substantial 
crystal electric field (CEF) that splits the ($2J+1$)-fold degeneracy 
of the spin-orbit-entangled total moment ${\bf J}$. 
For a half-integer moment $J$, the CEF ground 
state is a Kramers' doublet whose degeneracy is protected by
the time reversal symmetry. Often, the CEF gap is much larger than 
the temperature scale and exchange interaction in the system, and 
the low-temperature magnetic properties are fully captured by the 
ground state doublets that are modeled by pseudospin-1/2 local
moments.  

For the rare-earth double perovskites (Ba$_2$LnSbO$_6$,
Ln $=$ rare earths)~\cite{Shumpei2015,Dutta201664,PhysRevB.81.224409,
PhysRevB.68.134410,PhysRevB.65.144413,PhysRevLett.99.016404}, 
we propose a generic model on the FCC lattice 
that describes the nearest-neighbor interaction between the Kramers' 
doublet local moments. This generic model involves the Heisenberg 
interaction, the Kitaev interaction, and an additional crossing 
exchange that is symmetric in two pseudospin components. 
In the mean-field phase diagram of this generic model, 
we find large parameter regions that support 
ground states with continuous degeneracies.
Due to the spin-orbit entanglement, the generic model does not have 
any continuous symmetry. The continuous degeneracy is thus accidental 
and not related to any microscopic symmetry of the model. 
We expect that, the quantum fluctuations should break the accidental 
degeneracy and favor magnetic ordered states.  
This mechanism is known as order by quantum disorder 
(ObQD)~\cite{Henley1987,Villain1980,Savary2013,PhysRevLett.109.077204}.
Because of the continuous degeneracy, the fluctuations within the 
degenerate mean-field ground state manifold are very 
soft. Quantum fluctuations in a systematic $1/S$ expansion would 
lead to a small gap and a pseudo-Goldstone mode for large $S$, 
leading to a regime of temperatures with an additional magnetic 
contribution to the specific heat, $C_{\text{mag}} \sim T^3$. 
The impact of large quantum fluctuations for $S=1/2$ 
may further enhance the ObD gap; there is no controlled theory 
in this regime. In addition to the pseudo-Goldstone mode, 
the Weyl magnon mode~\cite{Feiye2016} is found
in the magnetic excitation for certain magnetic order. 
In contrast to the low energy pseudo-Goldstone mode, 
the Weyl magnon mode appears at finite energies due 
to the bosonic nature of the spin wave excitation. 

This paper is organized as follows. In Sec.~\ref{Sec: Ham}, we 
derive the generalized Kitaev-Heisenberg model.
We present a systematic analysis of the mean-field phase diagram 
of this model in Sec.~\ref{Sec: Classical}. Competition 
between different interactions, together with the geometrical frustration, 
leads to a very rich phase diagram. Specifically,
among different phases, we focus on the regions with a
continuous $U(1)$ or $O(3)$ degeneracy, in Sec.~\ref{Sec: Quan}. 
The degeneracy at the mean-field level is lifted when 
the quantum fluctuation is included, and various magnetic orders 
are favored in these regions. We demonstrate the ObQD explicitly. 
We further show the magnetic excitations of the resultant ordered phases
are characterized by the pseudo-Goldstone mode with a nearly gapless 
dispersion. Finally, we conclude with a discussion in 
Sec.~\ref{Sec: Discussion}.

\section{The generalized Kitaev-Heisenberg model}
\label{Sec: Ham}

We focus on a series of double perovskite-type oxides~\cite{Shumpei2015}, 
Ba$_2$LnSbO$_6$ (Ln$=$ rare earth), where 
the Ba ions are located at the A sites of the perovskite-type 
oxides ABO$_3$, and the Ln and Sb ions are regularly ordered at the B sites. 
Specifically, the Ln and Sb ions are ordered in the rock-salt type structure, 
with space group Fm$\bar{3}$m. Each of the two kinds of ions forms a 
separate FCC lattice. The magnetic behavior depends on the Ln$^{3+}$ ions  
([Xe]$4f^n$, [Xe]: electronic xenon core), where the 
SOCs are typically quite large. We study the Kramers' doublet 
that is formed by the $4f$ electrons of the Ln$^{3+}$ ion 
with an odd $n$ when the crystal electric field enters.

\begin{figure}
\includegraphics[width=0.36\textwidth]{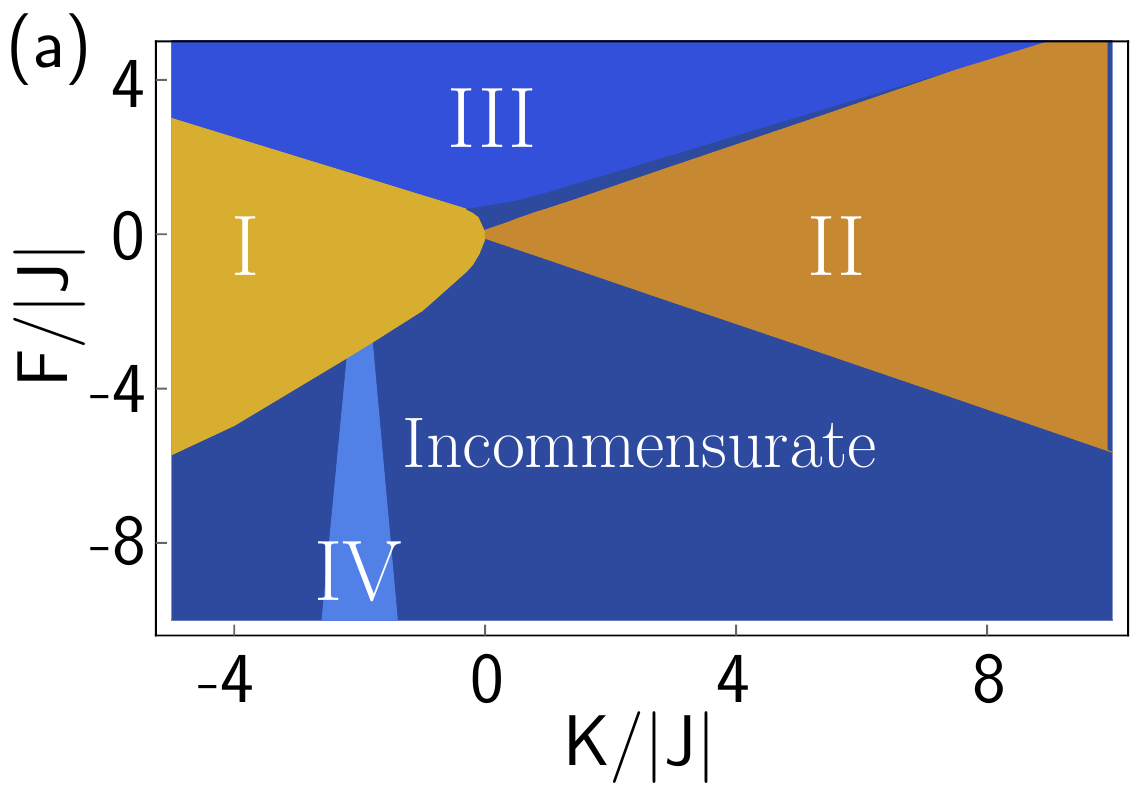}
\includegraphics[width=0.36\textwidth]{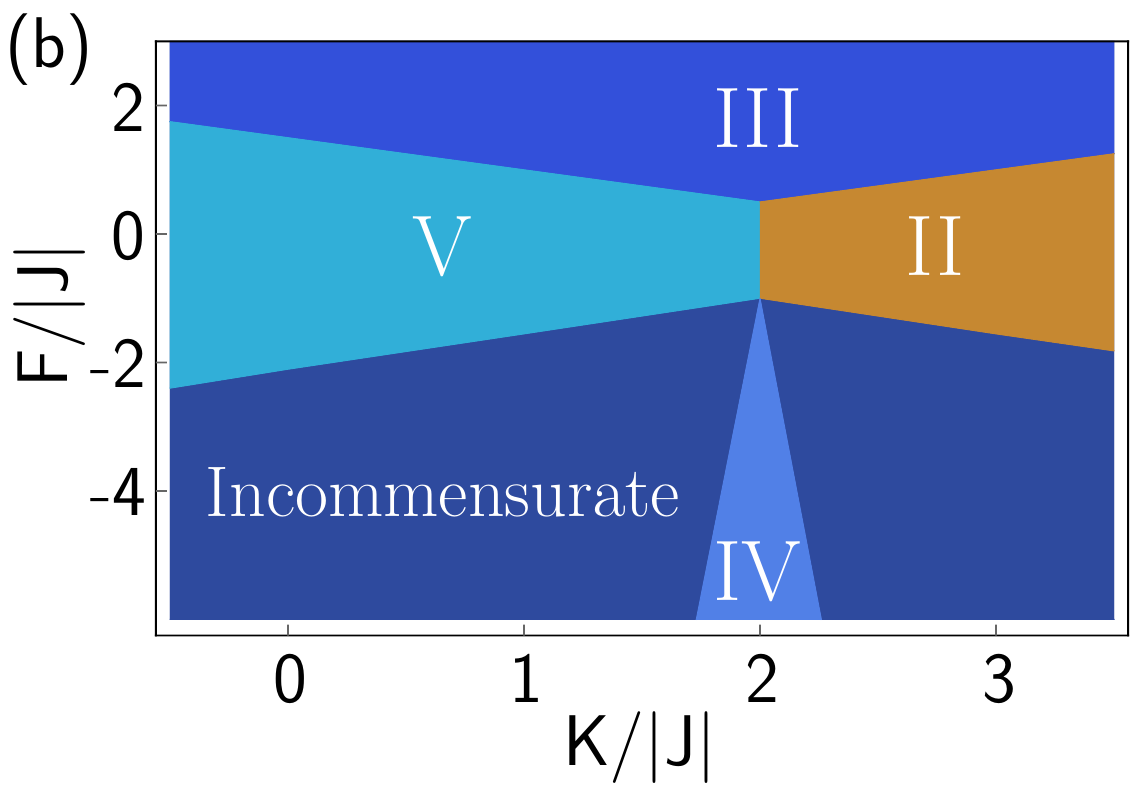}
\caption{(Color online.) The mean-field phase diagrams for
an antiferromagnetic Heisenberg coupling (a) and 
for a ferromagnetic Heisenberg coupling (b). 
The incommensurate phase has non-uniform spin amplitudes on every site. 
Both phase I and phase II have antiferromagnetic collinear orders 
with the wavevector X, and a continuous $U(1)$ ground state degeneracy 
exists in phase II. Both phase III and phase IV have antiferromagnetic 
collinear orders with the wavevector L, and phase IV shows a $U(1)$ degeneracy.
Phase V is ferromagnetically ordered with an $O(3)$ ground state degeneracy. 
See the main text and Tab.~\ref{table1} for a detailed discussion.
The lattice constant of the FCC lattice is set to unity throughout
the paper. 
}
\label{Fig2}
\end{figure}

Under the Fm$\bar{3}$m space group symmetry, the pseudospin,
$\bf {S}$, that acts on the Kramers' doublet of the rare earth ion, 
transforms as a pseudovector. Both the pseudospin position and the 
pseudospin orientation are transformed. The most general exchange 
interaction between the local moments on the nearest neighbor sites, 
allowed by the lattice symmetry, is a generalized Kitaev-Heisenberg 
model with 
\begin{equation}
H = \sum_{\vev{ij}_{\gamma \pm}} 
\big[ J\, {\bf S}_i^{}\cdot{\bf S}_j^{}
+ K S^{\gamma}_i S^{\gamma}_j 
\pm F ( S^{\alpha}_i S^{\beta}_j 
+ S^{\beta}_i S^{\alpha}_j) \big],
\label{eq: ham}
\end{equation}
where the bond index $\gamma \pm$ refers to the specific 
interaction that depends on the orientation of the bond 
in the plane and the pseudospin components are defined in the 
global coordinate system (see Fig.~\ref{Fig1}). 
We expect the nearest neighbor interaction is sufficient 
to describe the magnetic properties of the rare-earth moments 
in this system as the $4f$ electrons are very localized 
spatially. Besides the ordinary isotropic Heisenberg exchange 
interaction, we have the well-known Kitaev exchange 
interaction as well as the symmetric pseudo-dipole interaction 
that depends on the bond orientation. In Eq.~\ref{eq: ham}, 
the antisymmetric Dzyaloshinskii-Moriya interaction is prohibited 
by the inversion symmetry of the system~\cite{PhysRev.120.91}. 
The component $\gamma$ $(=x, y, z)$ specifies the three distinct 
types of Ising coupling in the Kitaev exchange ($K$ term), 
and $\{\alpha,\beta,\gamma\}$ is a cyclic permutation 
of $\{x, y, z\}$, that contributes to the symmetric 
pseudo-dipole interaction ($F$ term). The bond dependent 
pseudospin interaction is a direct consequence of the spin-orbit 
entanglement and widely occurs in many strong spin-orbit-coupled 
materials~\cite{Chen2008,Jackeli2009,Chen2010,Chen2011,li2016anisotropic}.

This generalized Kitaev-Heisenberg model was obtained previously 
by one of the authors and his collaborators in the context 
of the iridium-based double perovskites La$_2$BIrO$_6$ (B = Mg,Zn)~\cite{Cook2015,Aczel2016}. 
In the previous works, the mean-field phase diagram in the antiferromagnetic Heisenberg
regime was obtained with classical mean-field theory and classical Monte 
Carlo~\cite{Cook2015}, and the spin-wave  
spectrum was compared to the experiments in the regime with a dominant Kitaev 
interaction and a sizeable second-neighbor ferromagnetic interactions between 
the iridium local moments~\cite{Aczel2016}. Here, our motivation and purpose 
in this paper is different. We are inspired by the magnetic properties of 
the rare-earth double perovskites that host $4f$ electrons. 
As we have explained in Sec.~\ref{sec1}, the exchange interaction 
of $4f$ local moments is short-ranged and we only keep the nearest-neighbor 
interactions. This clearly differs from iridates. 
For iridates, there are five electrons (or one hole) in the triply degenerate 
$t_{2g}$ orbitals for the magnetic ion Ir$^{4+}$ in the cubic crystal field 
environment. The atomic spin-orbit coupling is active on the $t_{2g}$ orbitals
and entangles the atomic spin with the orbitals. The simplicity of the 
spin-orbit-entangled wavefunction and the Ir-O-Ir exchange path allows 
the determination of the exchange interaction from a microscopic perspective~\citep{Chen2008,Rau2014}. 
The Heisenberg part of the exchange interaction for iridates is often 
antiferromagnetic, and that is the reason that the previous work on the iridate
double perovskites~\cite{Cook2015} studied this regime. 
Because of the spatial extension of the Ir 5d electrons, the previous work
on iridates further explored the second neighbor exchange interaction~\cite{Aczel2016}. 
In contrast, the wavefunction of the Kramers doublet for the rare-earth moments
arises from the combined effect of the atomic spin-orbit coupling and 
the crystal electric field and depends sensitively on the crystal electric field
Hamiltonian that acts on the spin-orbit-entangled total moments. 
As a result, the exchange interaction between the rare-earth local moments 
varies for different wavefunctions of the Kramers doublets. Moreover, 
from the experience on rare-earth pyrochlore materials, many different parameter
regimes can occur~\cite{Savary2013,PhysRevX.1.021002}. 
So we explore all parameter range for the rare-earth double 
perovskites. In particular, while previous work on the 
iridates~\cite{Cook2015,Aczel2016} only studied the case of 
antiferromagnetic first neighbor coupling $J>0$, 
here we also allow a ferromagnetic $J<0$. In the remaining sections, 
we study this generalized Kitaev-Heisenberg model in all parameter 
regimes and obtain the magnetic properties and the magnetic excitations.

Compared with the rare-earth triangular system~\cite{li2016anisotropic,li2015rare,shen2016spinon} 
and the pyrochlore system~\cite{Onoda2011,Huang2014,Curnoe2008}, 
there are only three independent pseudospin interactions in Eq.~\ref{eq: ham}. 
It is the symmetries of the FCC lattice that help reduce the number 
of independent pseudospin interactions in our model. This result 
indicates that one may find even simpler models in strong 
spin-orbit-coupled systems with large lattice symmetries.

\section{Mean-field phase diagram}
\label{Sec: Classical}

\begin{table}
\begin{tabular}{cccc}
\hline  \hline \vspace{0.5mm}
Phase & Wavevector & Order Para. & Continuous deg  
\\
I   &  $(2\pi,0,0)$ & along [100] axis & -- \vspace{0.5mm}
\\
II  &  $(2\pi,0,0)$ & in (100) plane & $U(1)$ \vspace{0.5mm}
\\
III & $(\pi,\pi,\pi)$ & along [111] axis &  -- \vspace{0.5mm}
\\
IV  & $(\pi,\pi,\pi)$ & in (111) plane &  $U(1)$ \vspace{0.5mm}
\\
V   &   $(0,0,0)$ & any direction &  $O(3)$\vspace{0.5mm}
\\
\hline\hline
\end{tabular}
\caption{The mean-field phases in Fig.~\ref{Fig2}. 
The incommensurate phase is not included here.}
\label{table1}
\end{table}

We now discuss the mean-field phase diagram of the 
generalized Kitaev-Heisenberg model in Eq.~\ref{eq: ham}.
We systematically analyze the mean-field ground states 
in different parameter regimes. We consider both 
antiferromagnetic and ferromagnetic Heisenberg 
interactions with $J > 0$ and $J < 0$, respectively.

In the classical mean-field theory, we first treat the pseudospin 
as a classical vector that satisfies the hard constraint 
$|{\bf S}_i| = S$. The classical (mean-field) energy of the system
needs to be optimized under this local constraint on every lattice
site. This procedure is difficult as the local hard constraint is 
hard to implement. Instead, we here adopt the well-known Luttinger-Tisza 
method~\cite{PhysRev.70.954} that is to replace the local hard spin 
constraint by a global one such that 
\begin{eqnarray}
\sum_{i} |{\bf S}_i|^2 = N S^2, 
\end{eqnarray}
where $N$ is the total number of the pseudospins in the system. 
We optimize the classical mean-field energy, 
\begin{equation}
E_{\text{cl}} = 
\sum_{\bf q} \sum_{\alpha\beta} 
    \mathcal{E}_{\alpha\beta} ({\bf q}) 
    S_{\bf q}^{\alpha} \, S_{-{\bf q}}^{\beta},
\end{equation}
under the global constraint. Here we have defined 
\begin{equation}
S^{\alpha}_{i} = \frac{1}{N^{ \frac{1}{2}  }} 
\sum_{\bf q} S^{\alpha}_{\bf q}\, e^{i \, {\bf q}\cdot {\bf r}_i}.
\end{equation}
Once the mean-field ground state satisfies both the global
constraint and the local hard spin constraint, then the   
ground state under this approximation turns out to be the 
real ground state of the model in the classical limit.

In Fig.~\ref{Fig2}, we depict the mean-field phase diagram
with both antiferromagnetic and ferromagnetic Heisenberg 
interactions. The antiferromagnetic case with $J>0$ has a 
large classical degeneracy of ordered states, so that even 
a small Kitaev and $F$ coupling can influence the nature 
of the ground state in the vicinity of the pure Heisenberg point. 
However, for the ferromagnetic case with $J<0$, the 
anisotropic couplings primarily lead to pinning of the moments 
for small strengths, but large anisotropic couplings can 
change the nature of the ordered state. In the phase diagram, 
there is a large region where the minimum of the mean-field 
energy occurs in a set of incommensurate wavevectors (see Fig.~\ref{Fig2}). 
In these incommensurate regions, only one spin component 
is involved in the mean-field ground state. As a result, 
this incommensurate state cannot satisfy the local hard spin 
constraint due to the incommensurability. This result 
indicates the strong frustration in these regions of 
the generalized Kitaev-Heisenberg model. 
It is noted that the mean-field phase diagram in the 
regime with $-2<F/J<2, -2<K/J<2$ and $J>0$ 
is obtained in the previous work~\cite{Cook2015}. 

We continue with other ordered phases in the phase diagram.
In Fig.~\ref{Fig2}a, phase I is an antiferromagnetic state 
with the ordering wavevector at $\text{X}= (2\pi,0,0)$  
or equivalently $(0,2\pi,0)$, $(0,0,2\pi)$.
In this state, the spins order in a collinear pattern. 
For the $(2\pi,0,0)$ ordering wavevector,
the spin ordering is locked to the $\hat{x}$ direction with,
\begin{equation}
\text{I:} \quad\quad {\bf S}_i \equiv S \, \hat{m}_i = S \, \hat{x} 
 \, e^{2 \pi x_i },
\end{equation}
where $x_i$ is the $x$ coordinate of the lattice site ${\bf r}_i$. 
The locking between the ordering wavevector and the spin orientation
is a direct consequence and general phenomenon 
of the strong spin-orbit-coupled magnets.

In phase II with a dominant and antiferromagnetic Kitaev 
interaction ($K>0$), the system also orders with the wavevector X 
and equivalent ones.  Although having the same ordering wavevector, 
the ground state of phase II has a continuous $U(1)$ degeneracy. 
If we choose the $(2\pi,0,0)$ ordering wavevector, the ground state 
is parameterized as 
\begin{eqnarray}
\text{II:} \quad\quad {\bf S}_i \equiv S \, \hat{m}_i 
= S \, [ \cos \theta \, \hat{y} + \sin \theta\, \hat{z} ] \, e^{2 \pi x_i },
\label{Eq: U1_1}
\end{eqnarray}
where $\theta$ is an angular variable. 
This $U(1)$ degeneracy can be well understood, 
because the classical energy gained from the antiferromagnetic $K$ term
remains invariant when spin vectors are rotated within the $U(1)$ manifold.
Here the presence of a weak pseudo-dipole interaction does not lift the degeneracy. 
At the mean field level,
phase I and phase II are understood as the easy axis
along the [100] direction and easy plane anisotropy in the (100) plane
for the order parameter, respectively. Note that in region II there also exists a 
line degeneracy from X to W in the reciprocal space. 
Since only one spin component is involved, therefore, however, 
it can not form a normalized spin spiral order.

In the regimes dominated by the pseudo-dipole interaction ($F$ term), 
we obtain two other ordered phases. Phase III is an antiferromagnetic 
ordered phase with the ordering wavevector L $=(\pi,\pi,\pi)$ 
or equivalent ones. Given L $=(\pi,\pi,\pi)$, the spin ordering 
is locked to the $[111]$ direction with
\begin{eqnarray}
\text{III:}\quad\quad 
{\bf S}_i \equiv S \, \hat{m}_i = \frac{S}{\sqrt{3}} 
            (\hat{x} + \hat{y} + \hat{z})\, 
            e^{i \pi (x_i + y_i + z_i )} . 
\end{eqnarray}
Finally, phase IV has the same ordering wavevector as phase III but
has a $U(1)$ ground state degeneracy. 
For the $(\pi,\pi,\pi)$ ordering, the spin vector is parameterized as 
\be
\text{IV:}\quad {\bf S}_i \equiv S \hat{m}_i 
= S (\cos \theta \, \hat{u}_1 + \sin \theta\, \hat{u}_2) 
e^{i \pi (x_i + y_i + z_i )},
\label{Eq: U1_2}
\ee
where $\theta$ is an angular variable, and $\hat{u}_1$, $\hat{u}_2$ 
are two unit vectors in the (111) plane, chosen as $\hat{u}_1=
[1\bar{1}0]/\sqrt{2}$, $\hat{u}_2=[11\bar{2}]/\sqrt{6}$.
At the mean field level, phase III and Phase IV can be understood as the easy axis
along the [111] direction and easy plane anisotropy in the (111) plane
of the order parameter, respectively. Furthermore, 
like the case in phase II, a line degeneracy exists in the reciprocal space, 
from L to another equivalent L (e.g. from $(\pi,\pi,\pi)$ to $(\pi,\pi,-\pi)$).
Since the spins do not have uniform magnitudes, they cannot be 
the ground states.

\begin{figure}
	\includegraphics[height=.19\textwidth]{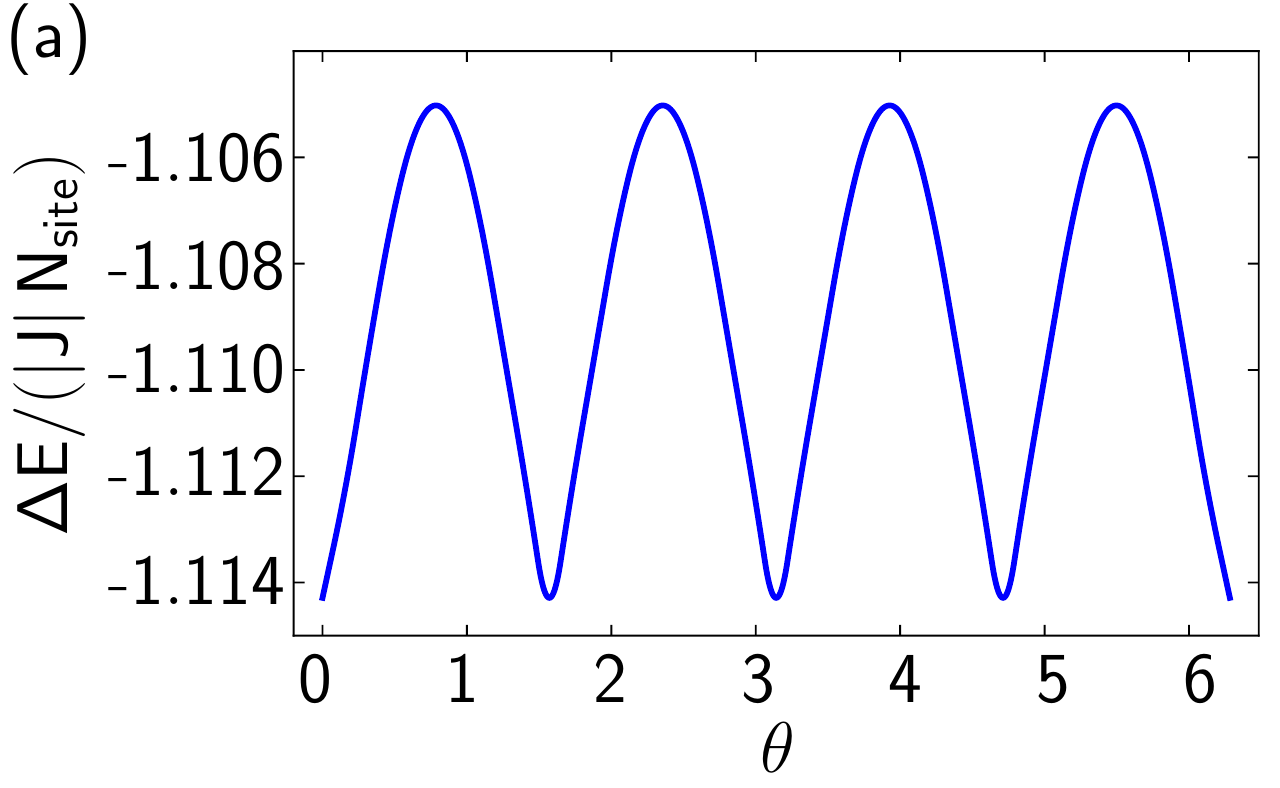}
	\includegraphics[height=.19\textwidth]{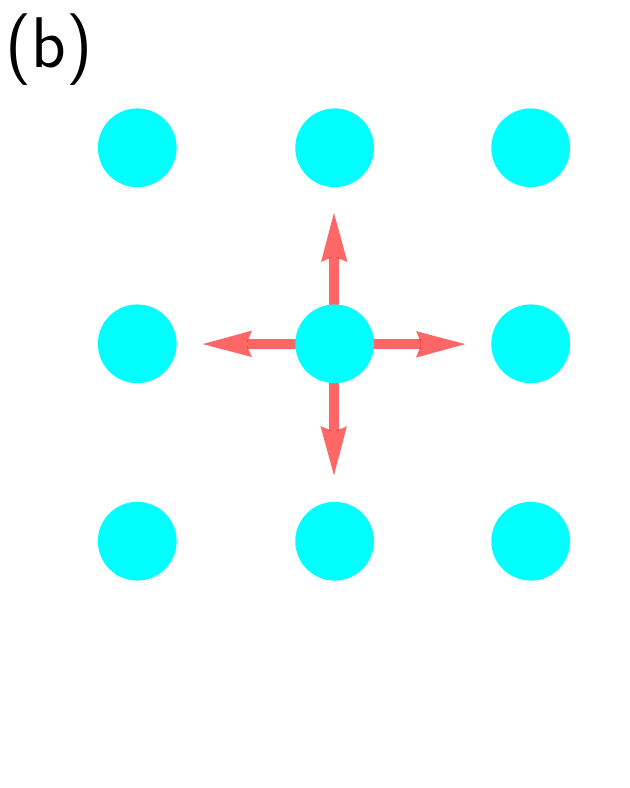}
	\caption{(Color online.) Quantum zero point energy in X $=(2\pi,0,0)$ ordered state (phase II). 
	(a) The minimum of $\Delta E$ occurs at $\theta = n\pi/2$ with $n \in \mathbb{Z}$.
	(b) Arrows indicate four-fold symmetry equivalent pseudospin orientation in the (100) plane. 
		We choose $(J,K,F)=(1,1,0)$.}
	\label{Fig3}
\end{figure}

When the Kitaev interaction is switched to ferromagnetic with $K<0$ 
and remains dominant, the ground state depends on 
the sign of the Heisenberg interaction. The case with an 
antiferromagnetic Heisenberg interaction gives phase I. 
For the ferromagnetic Heisenberg interaction with $J<0$, however, 
the classical ground state is a simple ferromagnetic state (phase V) 
but has an $O(3)$ degeneracy. The spin order is parametrized by 
two angular variables,
\begin{eqnarray}
\text{V:} \quad\quad
{\bf S}_i \equiv S \, \hat{m}_i &=& S \, (\sin \theta \cos \phi \, \hat{x} 
      + \sin \theta \sin \phi \, \hat{y} 
      \nonumber \\
&& \quad\quad \,\,+ \cos \theta \, \hat{z} ), 
\label{sphere}
\end{eqnarray}
where $\theta$ runs from 0 to $\pi$, 
and $\phi$ runs from 0 to $2\pi$. 
The $O(3)$ degeneracy, as in the previous $U(1)$ degeneracy case, 
is understood from the invariance of the dominant classical energy 
from the ferromagnetic Kitaev interaction.

As we summarize in Tab.~\ref{table1}, these five ordered phases have rather 
different order parameters. The phase transition between them, if there 
exists a direct transition between them, is first order.

\section{Quantum fluctuation and magnetic excitation}
\label{Sec: Quan}

We focus on the ordered phases with a continuous ground state degeneracy, 
and discuss the role of quantum fluctuation when the quantum nature of the
pseudospin is considered. Since the microscopic Hamiltonian only has discrete 
lattice symmetries, the continuous degeneracy of the mean-field ground states
is not granted at the quantum level. We, therefore, expect that the degeneracy 
in the mean-field level will be lifted when quantum fluctuation is included. 
Within the linear spin wave theory, we now discuss this order by quantum 
disorder (ObQD) effect explicitly.

\begin{figure}
\includegraphics[height=.19\textwidth]{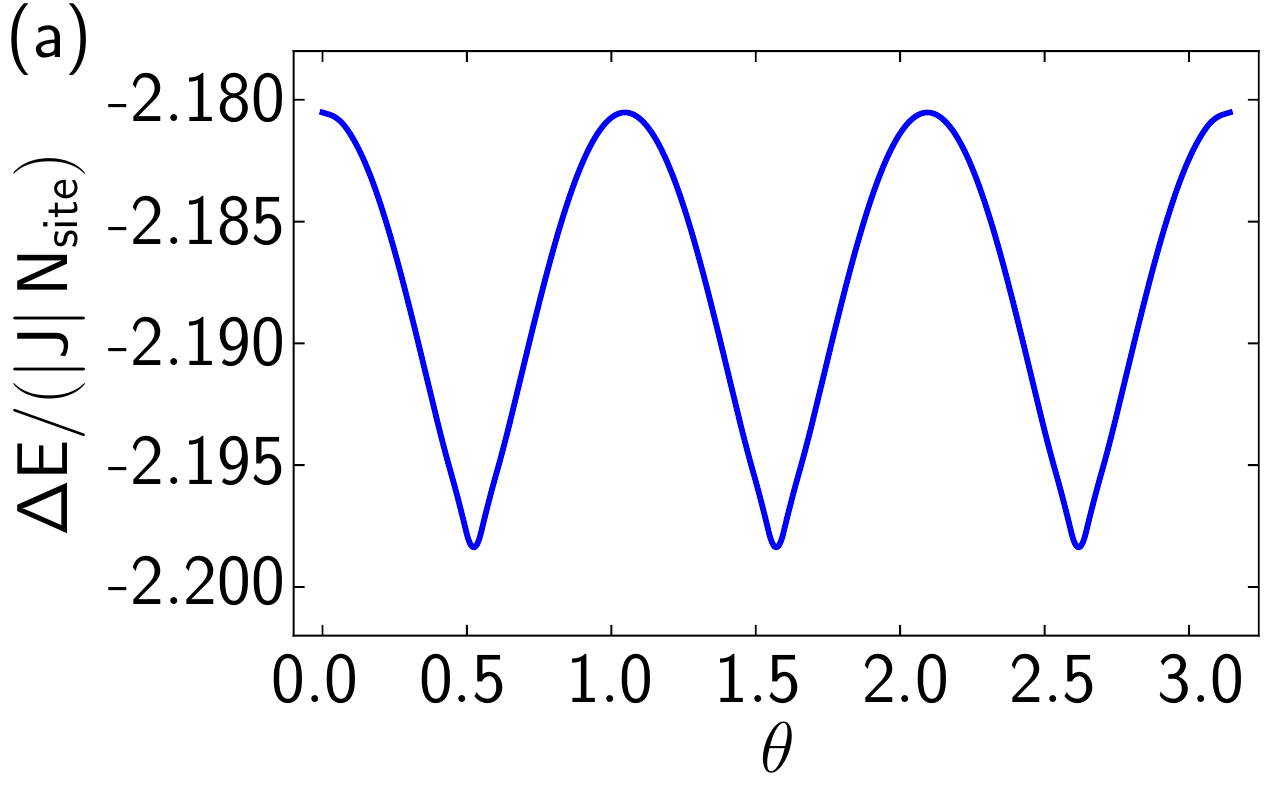}
\includegraphics[height=.19\textwidth]{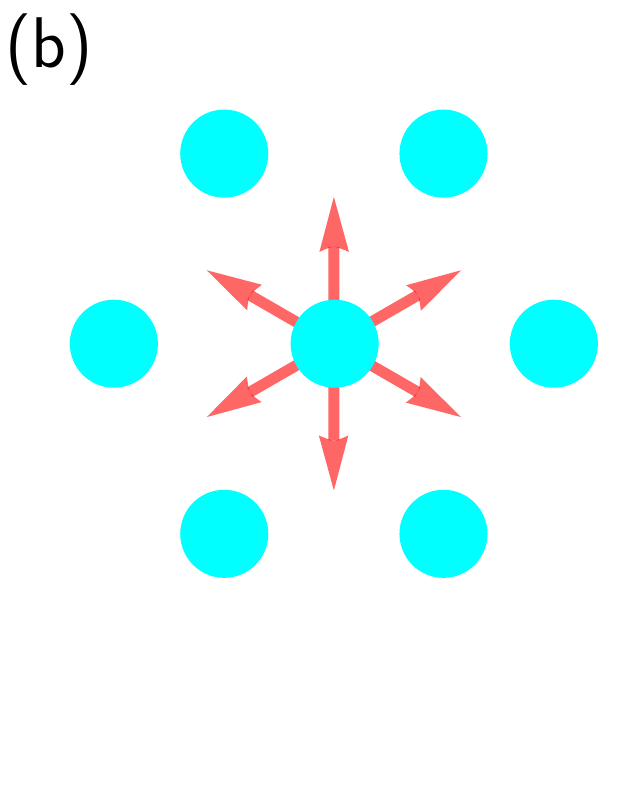}
\caption{(Color online.) Quantum zero point energy in L $=(\pi,\pi,\pi)$ 
ordered state (phase IV). (a) The minimum of $\Delta E$ occurs at 
$\theta = \pi/6 + n\pi/3$ with $n \in \mathbb{Z}$.
(b) Arrows indicate six-fold symmetry equivalent pseudospin orientation 
in the (111) plane. We choose $(J,K,F)=(-{1},2,-{4})$.}
\label{Fig4}
\end{figure}

For the ground state with a continuous $U(1)$ degeneracy, parametrized as in 
Eq.~\ref{Eq: U1_1} and Eq.~\ref{Eq: U1_2}, we introduce the Holstein-Primakoff 
bosons to express the spin operators as
\begin{eqnarray}
{\bf S}_i \cdot \hat{m}_i &=& S - b^\dagger_i b^{\phantom\dagger}_i,
\\
{\bf S}_i \cdot \hat{n}_i &=& \frac{(2S)^{ \frac{1}{2} }}{2} (b^{\phantom\dagger}_i + b^\dagger_i),
\\
{\bf S}_i \cdot (\hat{m}_i \times \hat{n}_i) &=& \frac{(2S)^{ \frac{1}{2} }}{2i}
(b^{\phantom\dagger}_i - b^\dagger_i), 
\end{eqnarray}
where $\hat{m}_i$ is the unit vector describing the spin orientation 
of classical spin order at site $i$, $\hat{n}_i$ is a unit vector 
normal to $\hat{m}_i$, and $S=1/2$. We substitute the spin operators 
with these Holstein-Primakoff bosons. In the linear spin wave 
approximation, we keep the boson terms up to the quadratic order. 
The resulting linear spin wave Hamiltonian has the following form 
\begin{eqnarray}
H_{\rm sw} &=& \sum_{\bf k} \Big[ \sum_{\mu, \nu} 
\big(
A^{}_{\mu\nu} ({\bf k}) b^\dagger_{{\bf k} \mu}
b^{\phantom\dagger}_{{\bf k} \nu} + B^{}_{\mu \nu}({\bf k})
b^{\phantom\dagger}_{-{\bf k}, \mu} 
b^{\phantom\dagger}_{{\bf k} \nu} 
\nonumber \\
&& \quad\quad + B^{\ast}_{\mu \nu}(-{\bf k}) 
b^\dagger_{{\bf k} \mu} b^\dagger_{-{\bf k}, \nu} 
\big) 
+ C({\bf k}) \Big] + E_{\rm cl},
\end{eqnarray} 
where $E_{\rm cl}$ is the classical mean-field energy of the ground state and 
independent of the angular variable $\theta$ due to the $U(1)$ degeneracy, 
$A_{\mu\nu}$, $B_{\mu\nu}$ and $C$ depend on $\theta$, 
and $A_{\mu\nu}$, $B_{\mu\nu}$ satisfy
\begin{eqnarray}
A_{\mu\nu}({\bf k}) &=& A^{\ast}_{\nu\mu}({\bf k}), \\ 
B_{\mu\nu}({\bf k}) &=& B_{\nu\mu}(-{\bf k}). 
\end{eqnarray}

While the classical mean-field energy $E_{\rm cl}$ preserves 
the $U(1)$ degeneracy, the quantum fluctuation lifts this 
continuous degeneracy through the quantum zero point energy
$\Delta E$ that is given by  
\be
\Delta E = \sum_{\bf k} \Big[  \sum_\mu \frac{1}{2} 
\big( \omega_\mu({\bf k}) - A_{\mu\mu}({\bf k}) \big) 
    + C({\bf k})  \Big],
\ee
where $\omega_\mu({\bf k})$ is the excitation energy of the $\mu$-th
spin wave mode at momentum ${\bf k}$.

\begin{figure}[t]
    {
        \includegraphics[height=.19\textwidth]{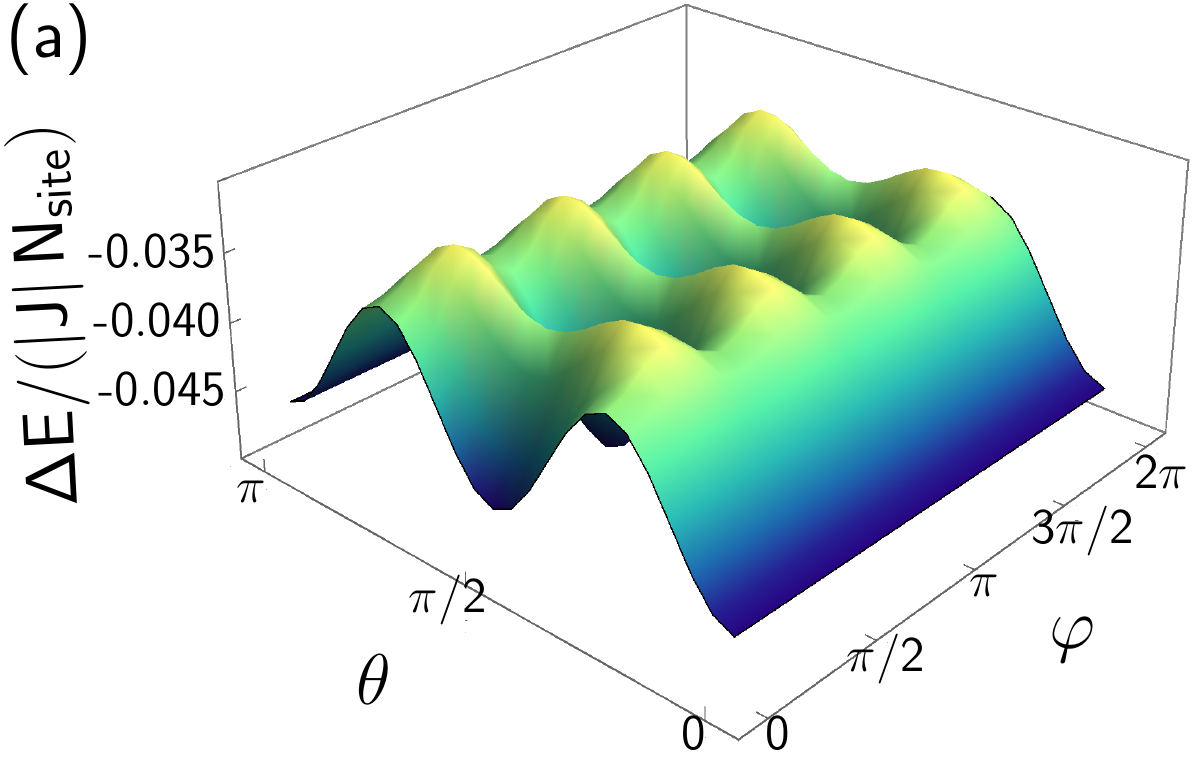}
        \includegraphics[height=.19\textwidth]{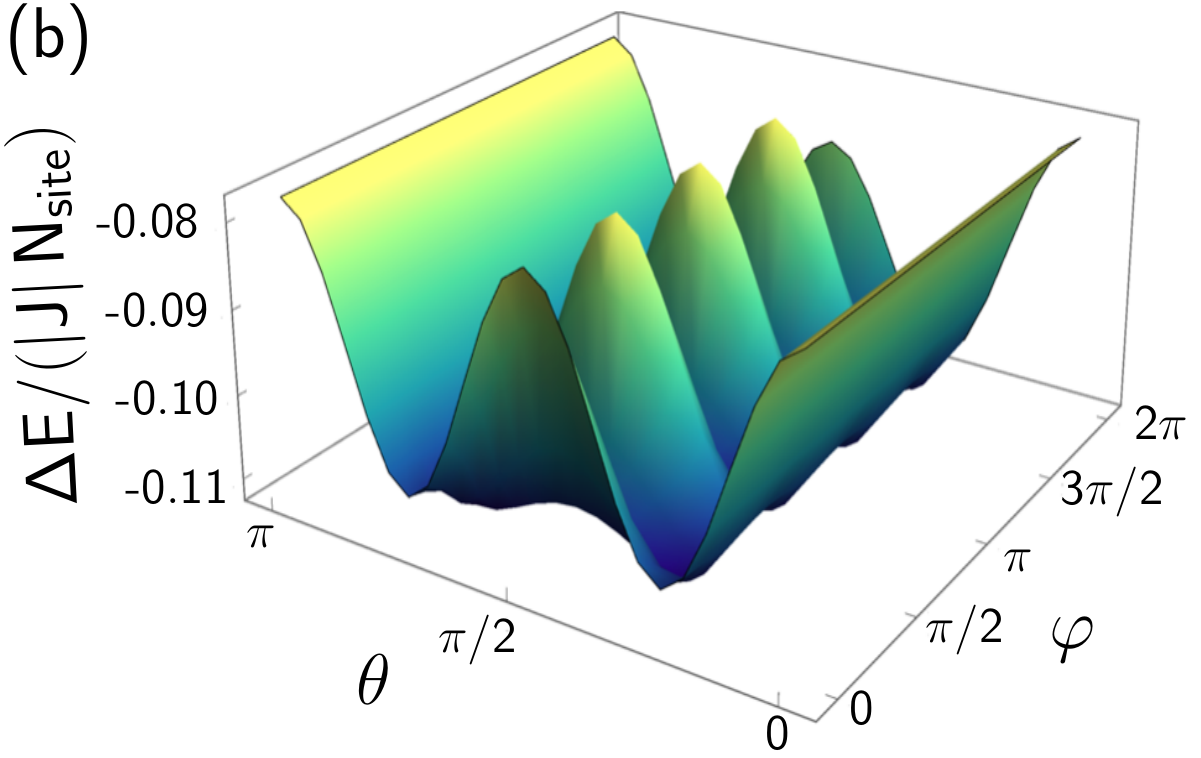}
    }
    \caption{(Color online.) Quantum zero point energy of two representative parameters in 
    mean-field phase V.
        The two-dimensional sphere is parametrized by polar angle 
        $\theta$ and azimuthal angle $\varphi$. 
        (a) We set $(J, K, F) = (-{1}, 1, 0)$ for phase V$_{\text a}$ in Fig.~\ref{Fignew}, 
        the favored spin orientations are $\pm \hat{x}$, $\pm \hat{y}$, 
        and $\pm \hat{z}$, corresponding to the six minima of $\Delta E$.
        (b) We set $(J, K, F) = (-{1}, 0, 1)$ for phase V$_{\text b}$ in Fig.~\ref{Fignew}, 
        the favored spin orientations are [111] directions, corresponding to 
        the eight minima of $\Delta E$.
      }
    \label{Fig5}
\end{figure}

In phase II where there is a U(1) degeneracy (see Eq.~\ref{Eq: U1_1}), 
the minima of zero point energy $\Delta E$ occurs
 at  
\begin{equation}
\text{II:} \quad\quad  \theta =  \frac{n\pi}{2},
\end{equation} 
 where $n \in \mathbb{Z}$, and the favored magnetic order has 
a four-fold symmetry equivalent configuration that is shown in Fig.~\ref{Fig3}.
 
In phase IV where there is also U(1) degeneracy (see Eq.~\ref{Eq: U1_2}), 
the minima of zero point energy $\Delta E$ occurs at 
\begin{eqnarray}
\text{IV:} \quad\quad \theta = \frac{\pi}{6} + \frac{n \pi}{3},
\end{eqnarray}
 where $n \in \mathbb{Z}$ and the favored magnetic order has a six-fold symmetry 
equivalent configuration that is shown in Fig.~\ref{Fig4}.

\begin{figure}[t]
{ \includegraphics[width=.36\textwidth]{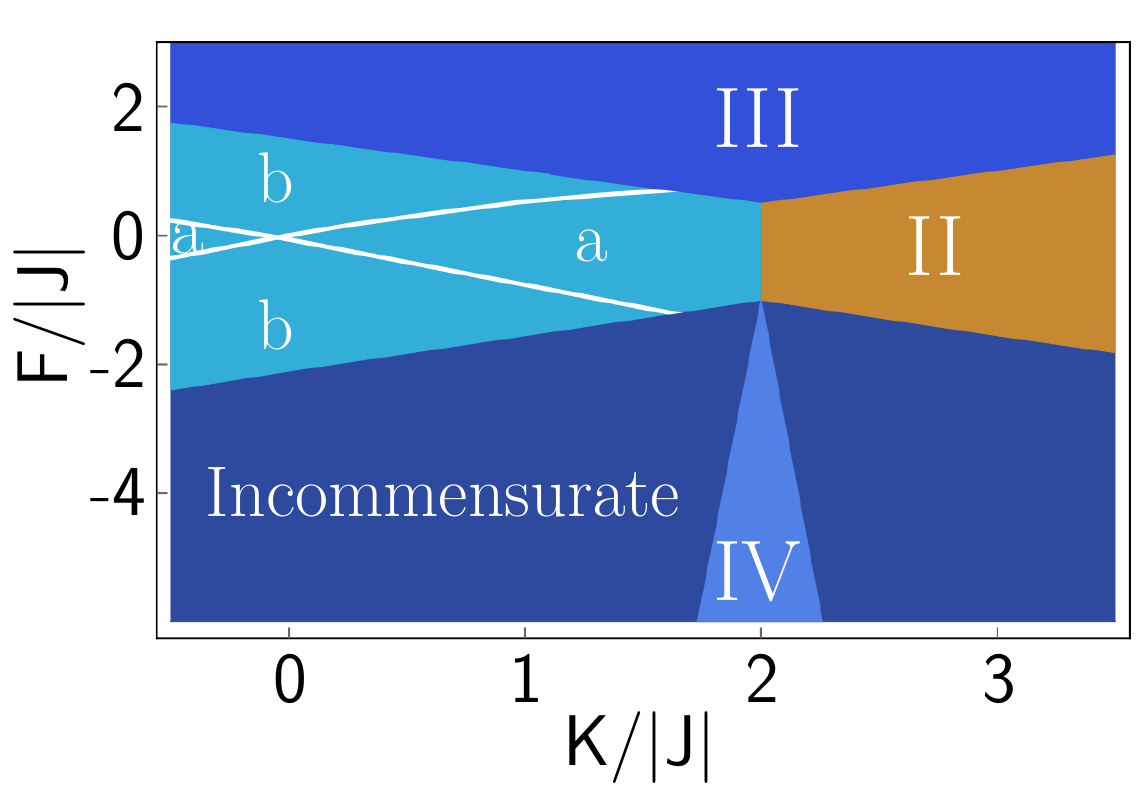} }
\caption{(Color online.) Phase V in the mean-field phase diagram 
of Fig.~\ref{Fig2} (b) is split into two phases with different 
magnetic orders once the quantum fluctuation is considered. Here $J<0$.}
\label{Fignew}
\end{figure}

Now we turn to phase V of the mean-field phase diagram, 
the ferromagnetic ordered state with an $O(3)$ 
degeneracy. As we have parametrized with a vector on a unit sphere in  
Eq.~\ref{sphere}, two angular variables ($\theta$ and $\phi$) are needed 
to capture the $O(3)$ degeneracy. The minima of the zero point energy 
$\Delta E$ are shown in Fig.~\ref{Fig5}. We find two distinct ordering 
patterns that are not equivalent under the lattice symmetry. 
As we depict in Fig.~\ref{Fignew}, the phase V of the mean-field
phase diagram is split into two distinct phases (V$_{\text a}$ 
and V$_{\text b}$). In V$_{\text a}$ (V$_{\text b}$), the 
quantum fluctuation selects the $[001]$ type ($[111]$ type) 
of magnetic order. 

The lifting of the $O(3)$ degeneracy is understood through a cubic anisotropy
that is induced by the quantum fluctuation. The cubic anisotropy in the 
energy is given as 
\begin{equation}
E_{\text{ani}} = \lambda_{\text{ani}} [(M^x)^4 + (M^y)^4 + (M^z)^4 ],
\end{equation}
where ${\bf M}$ is the order parameter of the ferromagnetic phase. 
In phase V$_{\text a}$, $\lambda_{\text{ani}}<0$ and we have the $[001]$ ordering.
In phase V$_{\text b}$, $\lambda_{\text{ani}}>0$ and we have the $[111]$ ordering.

\begin{figure}[tp]
{
\includegraphics[width=.23\textwidth]{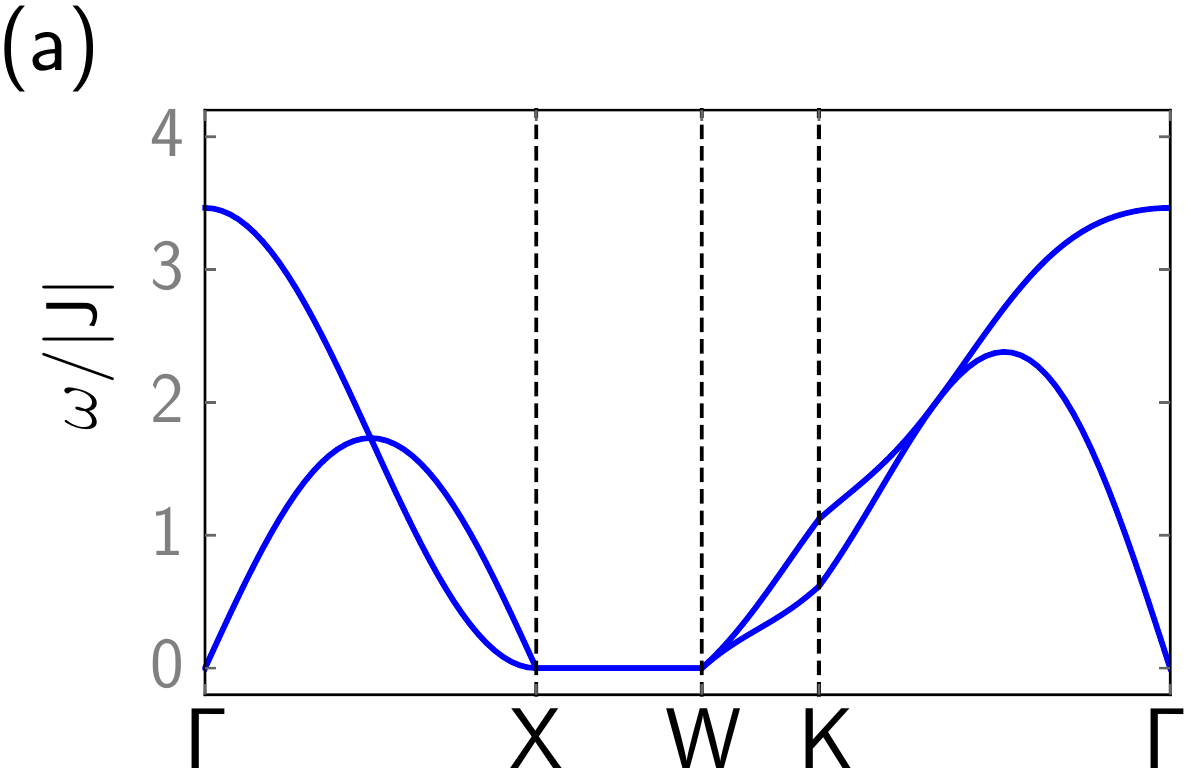}
\includegraphics[width=.23\textwidth]{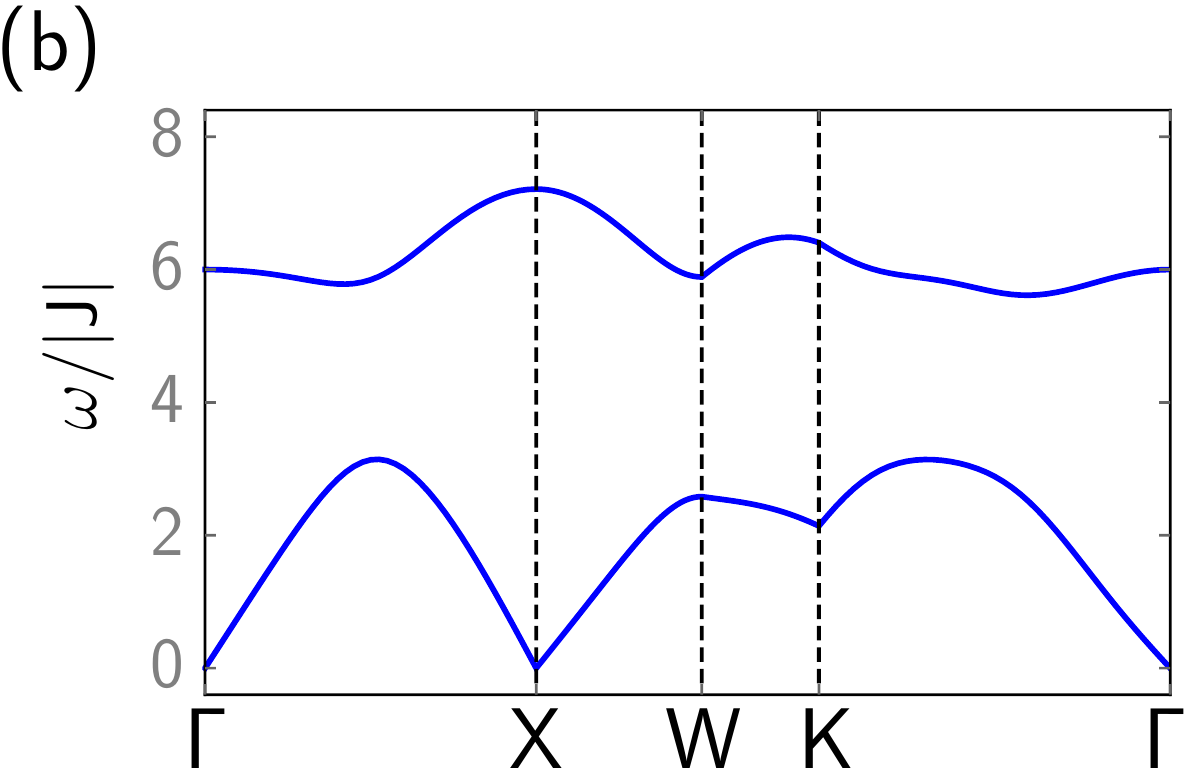}
\includegraphics[width=.23\textwidth]{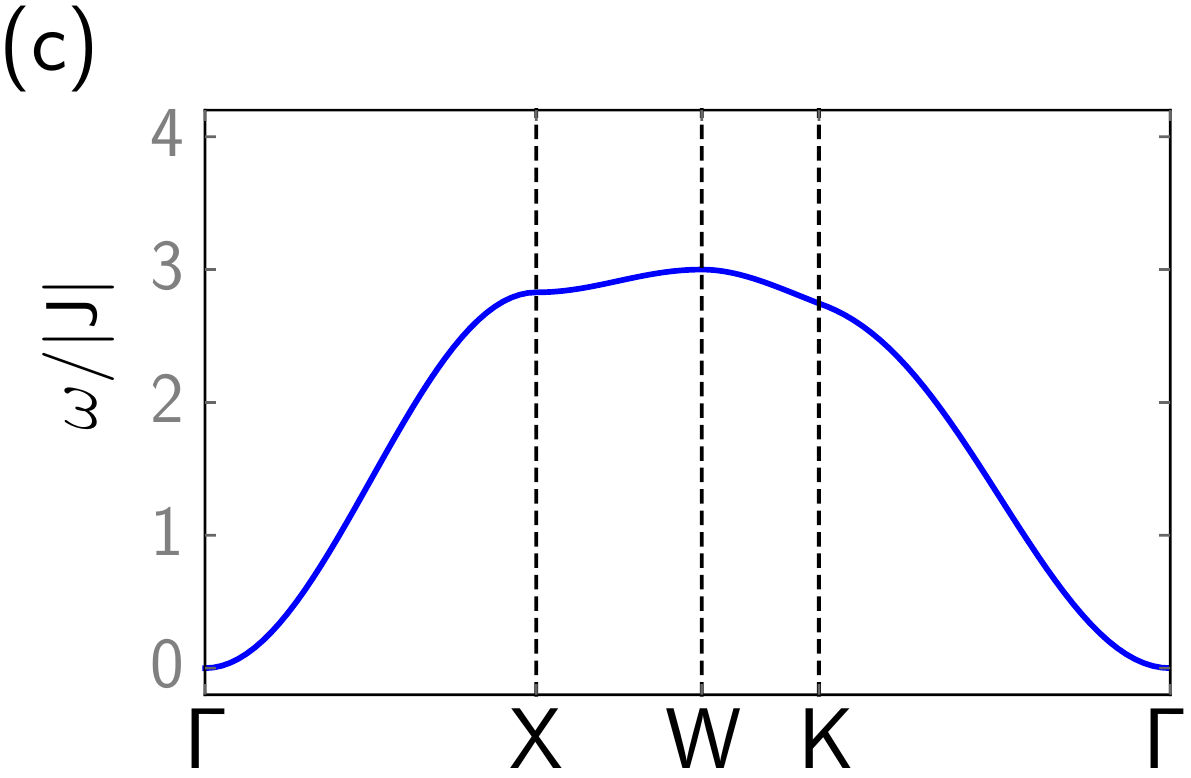}
\includegraphics[width=.23\textwidth]{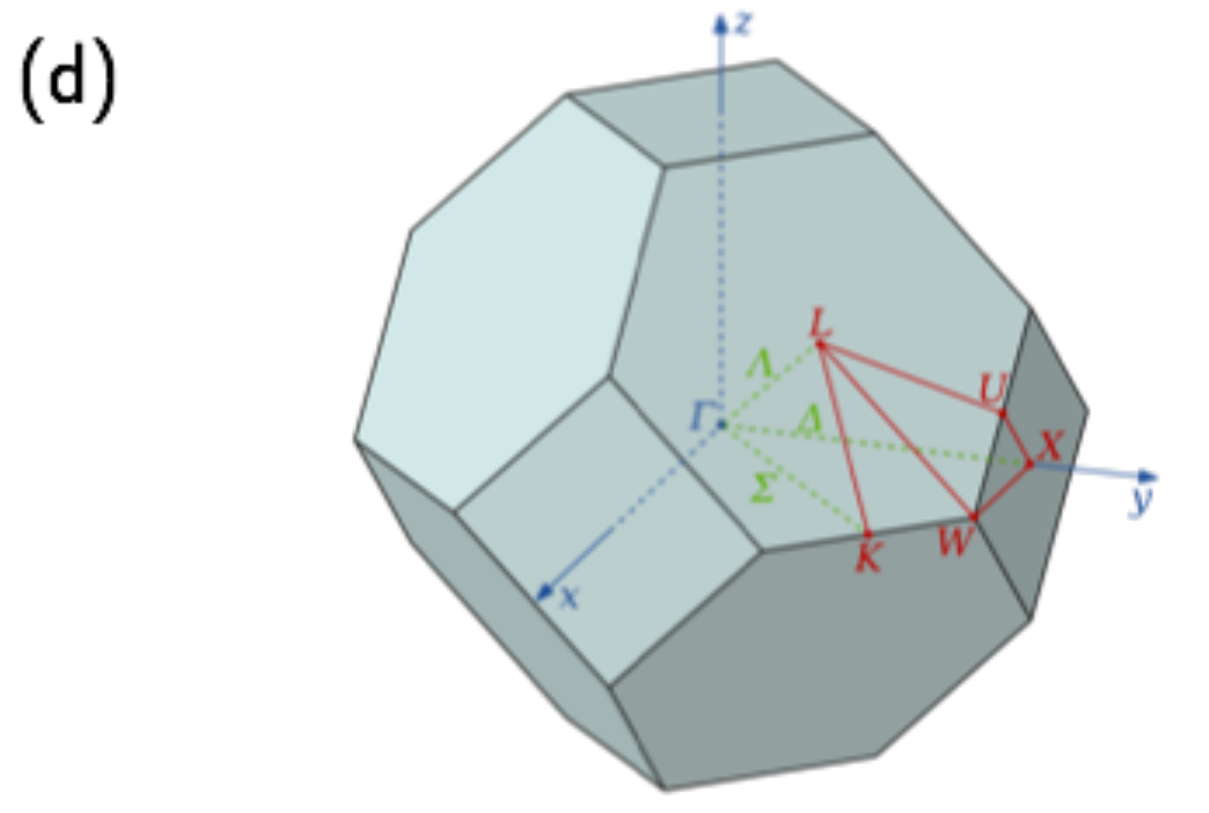}
}
 \caption{(Color online.)
 The representative spin wave spectra along high symmetry momentum lines with 
 (a) $(J,K,F) = (1,1,0)$ and $\theta = 0$ in phase II; 
 (b) $(J,K,F) = (-1,2,-4)$, $\theta = \pi/6$ in phase IV; 
 (c) $(J, K, F) = (-1, 1, 0)$, $\theta = \pi/2$, $\varphi = 0$ in phase V$_{\text a}$. 
 In (d) we depict the FCC Brillouin zone (the figure is adapted from Wikipedia~\cite{wiki}).}
\label{Fig6}
\end{figure}

Having determined the ground state configurations, 
we further study the spin wave excitation 
spectra in different phases. The results are depicted 
along high symmetry momentum lines in Fig.~\ref{Fig6}. 
There are two qualitative features in the spin wave 
spectra. First, we observe gapless modes in Fig.~\ref{Fig6} 
(a), (b) and (c). These pseudo-Goldstone modes are 
characteristic of the phase ordered due to quantum 
fluctuation that lifts the continuous degeneracy, 
in our case, phase II, IV and V. Although a gap is 
expected to be generated by anharmonic effects, 
nearly gapless dispersion is a possible experimental 
signature of the order by quantum disorder scenario.
Second, magnon spectrum in Fig.~\ref{Fig6} (a) shows 
band touching (along K-$\Gamma$) due to accidental degeneracy, 
indicating the ``Weyl magnon'' behavior and the corresponding 
topologically robust surface states, although the Weyl node 
along K-$\Gamma$ belong to the type-II
node~\cite{BernevigNature} for the specific parameter choice 
in Fig.~\ref{Fig6} (a). The Weyl band touching of the magnon
spectrum is expected to be stable even beyond the linear 
spin-wave theory due to the robust topological nature.

\section{Discussion}
\label{Sec: Discussion}

Despite the abundance of the rare-earth double perovskites~\cite{Shumpei2015,Dutta201664}, 
the experimental characterization of them is quite limited.
Only the crystal structure and the magnetic susceptibility 
measurements have been carried out so far. All these compounds 
are paramagnetic and have no magnetic ordering down to 1.8K~\cite{Shumpei2015}. 
This result does not mean all of them would be spin liquids. 
The temperature (1.8K) is not quite low for the rare-earth local
moments since the exchange interaction between them is of the 
order of a couple Kelvin. It is very likely that the absence of 
magnetic ordering down to 1.8K is a thermal effect, and 
one could observe the ground state properties if the temperature
is further lowered.

What would be the experimental phenomena that are expected for 
the ObQD phenomena and Weyl magnons? 
Clearly, one should observe one of the orders that we predict. 
Moreover, the consequence of the ObQD is the presence of the nearly
gapless pseudo-Goldstone mode for the magnetic excitation. 
Strictly speaking, the pseudo-Goldstone mode
should develop a minigap due to the anharmonic quantum effect, 
but one would expect a $T^3$ heat capacity in the temperature 
regime above the minigap energy scale. For the Weyl magnons, 
one could probe the spin wave spectrum with the inelastic 
neutron scattering measurement and directly detect the linear 
band touching. Alternatively, one could measure the consequence of 
Weyl magnons, such as the chiral surface state 
and optical conductivities. All these probes
have been discussed in details in Ref.~\onlinecite{Feiye2016}. 

In many systems, the ObQD is very fragile because 
other small interactions, that are not included in the model,
may simply drive the system into a different state. 
For rare-earth double perovskites, however, we expect the dominant
interaction is from the nearest neighbors. The further neighbor  
exchanges are rather weak due to the spatial localization of the 
$4f$ electrons. The remaining interaction is the magnetic dipole 
interaction that decays very fast with the separation of the local moments. 
The actual magnitude would depend on the material's details 
such as the the moment size and lattice constants. In any case, 
we expect the rare-earth double perovskites to be promising 
candidates for the ObQD phenomena. 

The study of the rare-earth double perovskites is in the early stage. 
Many physical properties of the rare-earth double perovskites
need to be measured, and it is very likely that other exotic 
quantum phases could emerge besides the ones that have been predicted 
here. We expect our work to bring further attention to this new class
of materials. 

\emph{Acknowledgements.}---This work is supported by the Ministry of 
Science and Technology of People's Republic of China with the Grant 
No. 2016YFA0301001 (G.C.), NSERC of Canada (A.P.), the Start-Up Funds 
of Fudan University (Shanghai, People's Republic of China) and the 
Thousand-Youth-Talent Program (G.C.) of People's Republic of China.
Research at Perimeter Institute is supported by the Government of 
Canada through the Department of Innovation, Science and Economic 
Development Canada and by the Province of Ontario through the 
Ministry of Research, Innovation and Science.

\bibliography{ref}

\end{document}